# Title

Modeling Dynamic Computations in the Primate Ventral Visual Stream

# Authors

Matteo Dunnhofer[1,2]*, Maren Wehrheim[1,3]*, Hamidreza Ramezanpour[1]*, Sabine Muzellec[1]*, and Kohitij Kar [1]

# Affiliation

1. York University, Department of Biology, Centre for Vision Research, Centre for Integrative and Applied Neuroscience, Toronto, Canada
2. University of Udine, Department of Mathematics, Computer Science, and Physics, Udine, Italy
3. Mila, the Quebec Artificial Intelligence Institute

\* denotes equal contribution
Correspondence should be addressed to Kohitij Kar
E-mail: k0h1t1j@yorku.ca

# Conflict of interest

The authors declare no competing financial interests.

# Acknowledgments

KK has been supported by funds from the Canada Research Chair Program, the Simons Foundation Autism Research Initiative (SFARI, 967073), Brain-Canada Foundation, the Canada First Research Excellence Funds (VISTA Program), and the National Sciences and Engineering Research Council of Canada (NSERC). MD received funding from the European Union's Horizon Europe research and innovation programme under the Marie Skłodowska-Curie grant agreement n. 101151834 (PRINNEVOT). MW and SM are funded by the Connected Minds Postdoctoral Fellowship (supported by CFREF).

# Abstract

A major goal of computational neuroscience has been to explain how the primate ventral visual stream (VVS) transforms visual input into temporally evolving neural representations that support robust visual perception. Historically, most modeling efforts have assumed static conditions: monkeys fixate a dot, images are briefly flashed, and neural responses are analyzed through time-averaged metrics. Feedforward deep networks trained on static object recognition tasks outperform prior work in approximating these static snapshot-driven VVS responses. However, mounting neurophysiological evidence demonstrates that VVS responses are rich dynamical signals shaped not only by the retinal input but also by intrinsic circuit dynamics, recurrent interactions, and widespread top-down modulation. Moreover, real-world vision is inherently dynamic: objects move, the observer moves, and the eyes actively sample the environment. Here, we review recent progress in modeling dynamic responses in the macaque ventral stream across three domains: (1) intrinsic dynamics elicited by static images, (2) dynamics evoked by dynamic visual stimuli, and (3) dynamics generated by active sensing during eye movements. We argue that accurately modeling VVS dynamics will require representational, circuit-level, and behavioral perspectives, including multi-area recurrence, structured E/I interactions, and temporal objectives that better reflect natural behavior. We outline some key missing ingredients and propose a roadmap toward dynamic, multi-timescale models of the primate VVS.

# 1. Introduction

## 1.1 The classic modeling problem in visual neuroscience

A significant portion of what we know about object recognition in the primate brain comes from a highly standardized paradigm (Figure 1A): a fixating monkey, briefly flashed images, and neural responses analyzed within fixed temporal windows [1–3]. Within this framework, representational geometry is typically quantified using distance-based metrics such as representational dissimilarity matrices (RDMs [4]) and centered kernel alignment (CKA [5]). In parallel, predictive modeling approaches have used linear mappings to relate image-evoked neural activity to features from artificial neural network (ANN) models trained on object categorization tasks [6–8], achieving strong correspondence particularly in early post-stimulus epochs [9]. These comparisons revealed that specific feedforward ANNs capture a large fraction of variance in the time-averaged structure of the ventral stream's responses across images, and currently stand as the best hypotheses for ventral stream computations. However, these successes hinge on analyses that largely ignore when and how neural representations evolve over time.

## 1.2 The problem: the ventral stream is inherently dynamical, the world is dynamic, and our sensing system is dynamic

The ventral visual stream (VVS) neurons express rich temporal structure even when images are briefly flashed [9–11]. These temporal dynamics arise from local recurrent circuitry, long-range feedback from higher-order areas, intrinsic fluctuations, and the animal's own behavioral state. Moreover, real-world vision is fundamentally dynamic: objects move, and eye movements continuously reshape retinal input. Thus, models trained on static images and evaluated with static metrics do not fully capture the computations performed by VVS neurons across time.

## 1.3 Goal of this review

In this review, we outline the major drivers of ventral stream dynamics, survey existing classes of dynamic computational models, and integrate findings from dynamic stimuli and active sensing. We end by arguing that current models, which focus on single-neuron responses or representational trajectories, capture only part of the story. Fully predicting VVS dynamics will require models that incorporate neuron–neuron interactions, structured E/I circuitry, and population-level synchrony—core circuit features that shape temporal computation.

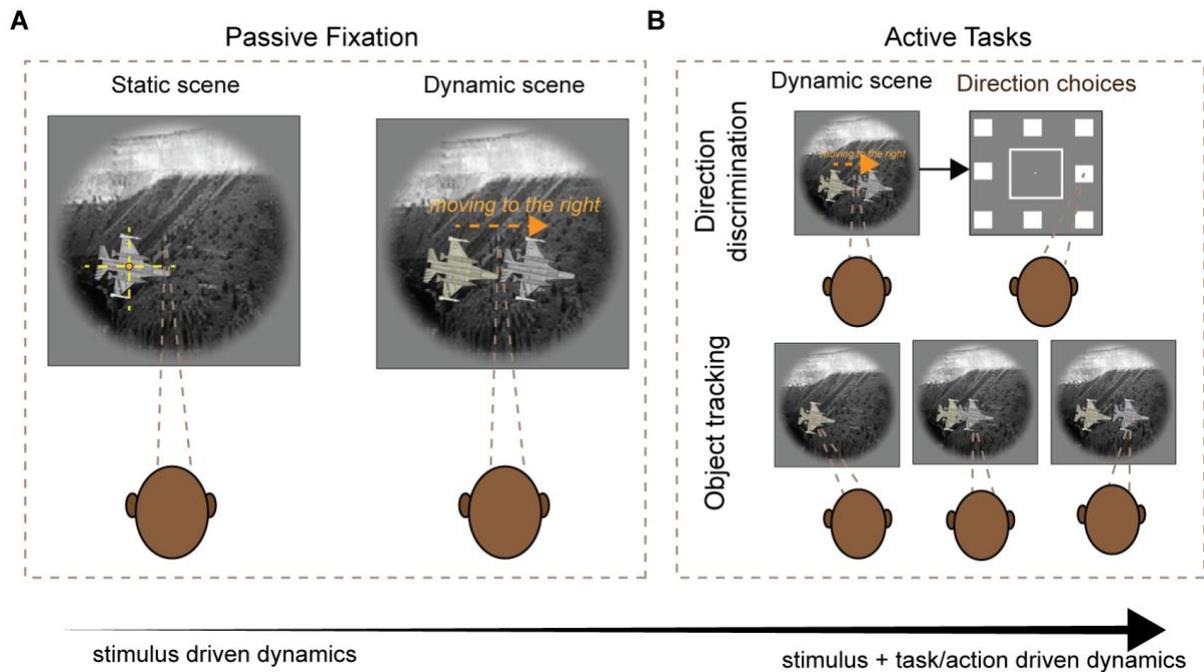

**Figure 1. Experimental regimes used to probe dynamic computation in the primate ventral visual stream. A. Passive fixation.** Monkeys maintain fixation while viewing either static scenes (left) or dynamic scenes (right). Even for static images, ventral-stream responses evolve over time due to intrinsic circuit dynamics and recurrent processing [9,12]. Dynamic scenes introduce temporally structured sensory input, enabling the study of stimulus-driven neural dynamics under controlled fixation. **B. Active tasks.** During direction discrimination (top), subjects actively report the object's motion direction, thereby engaging decision-making and top-down modulation. During object tracking (bottom), continuous eye movements and behavioral goals dynamically reshape retinal input across time. The horizontal axis highlights a progression from predominantly stimulus-driven dynamics to dynamics jointly shaped by stimulus structure, task demands, and action, motivating the need for models that capture multi-timescale visual computation under naturalistic conditions. Figured adapted from [13].

## 2. Neural Dynamics for Static Images: The VVS as a Dynamical System

### 2.1 Empirical observation: VVS responses to static images are temporally structured

Responses in the VVS to static images unfold over time in a structured, functionally meaningful manner, reflecting successive computational stages rather than merely decaying feedforward signals. While, on average, neural response latencies progress reliably along the ventral pathway from early visual cortex to the inferior temporal (IT) cortex (V1 → V2 → V4 → IT), early and late responses within the same cortical area often encode qualitatively different information. Classic work by Sugase et al. [10] demonstrated that early IT responses carry coarse category information (e.g., face vs. non-face), whereas later activity refines identity-level information. Time-resolved population decoding has since shown that object identity can be rapidly read out from IT but that it continues to evolve over time [1,14]. Kar et al. [9] showed that late-phase IT responses are selectively necessary for solving object identity for images that are not easily solved by feedforward ANN models. Relatedly, Shi et al. [11] observed a similar temporal code-switching in IT responses, where the informational content of population activity evolves from detection to fine-grained discrimination. Wehrheim et al. [15] observed a temporal dissociation between overall task accuracy and the behavioral alignment of neural decodes for facial expressions. These observations underscore how neural population codes continue to evolve and restructure over tens to hundreds of milliseconds, even in the absence of new visual input or additional eye movements, underscoring the need to understand how these dynamics are generated.

## 2.2 What brain areas generate these dynamics?

These temporal dynamics raise a mechanistic question: where do they come from? VVS dynamics arise from a combination of intrinsic local recurrence and top-down recurrent loops with other cortical and subcortical areas. Anatomically, the VVS is densely recurrent in addition to its feedforward organization: horizontal connections within V1, V2, and V4 link similarly tuned neurons over millimeters of cortex, and feedback projections from IT to V4 and from V4 to V2 are numerous and topographically organized [16–19]. At the microcircuit level, laminar-specific connectivity separates early feedforward drive from later recurrent and feedback influences, with feedforward input targeting layer 4 and feedback dominating supra- and infragranular layers [20,21]. Beyond intrinsic circuitry, late-phase visual activity is strongly shaped by top-down inputs: vlPFC provides delayed, behaviorally relevant feedback to IT [12], hippocampus and amygdala modulate IT according to memory and affect [22–24], and the pulvinar–parietal network gates and synchronizes VVS processing [25,26].

## 2.3 Where We Stand: Modeling Static-Image Dynamics

Together, the evidence shows that, even for static images, the VVS performs temporally multiplexed computations within a recurrent brain-wide network. Yet most vision models (for summary, see [27]) remain largely time-agnostic. Feedforward networks collapse processing into a single pass, allowing models with similar peak predictivity to implement very different temporal computations. Recurrent architectures improve alignment to late-phase neural responses and challenging-image behavior [28–32]. But without constraints on latencies, phase-specific coding, or multi-area interactions, they risk improving biological fidelity only marginally. Addressing this gap requires moving beyond static similarity measures such as RSA [4] or CKA [5] toward approaches that explicitly evaluate representational trajectories over time. Methods such as dynamic RSA [33] and Dynamical Similarity Analysis [34] provide principled ways to compare when and how representations evolve, highlighting that models must capture not only static states but the underlying dynamics that generate behaviorally relevant late representations. These findings raise a natural next question: how are ventral-stream dynamics further shaped when visual input itself unfolds over time?

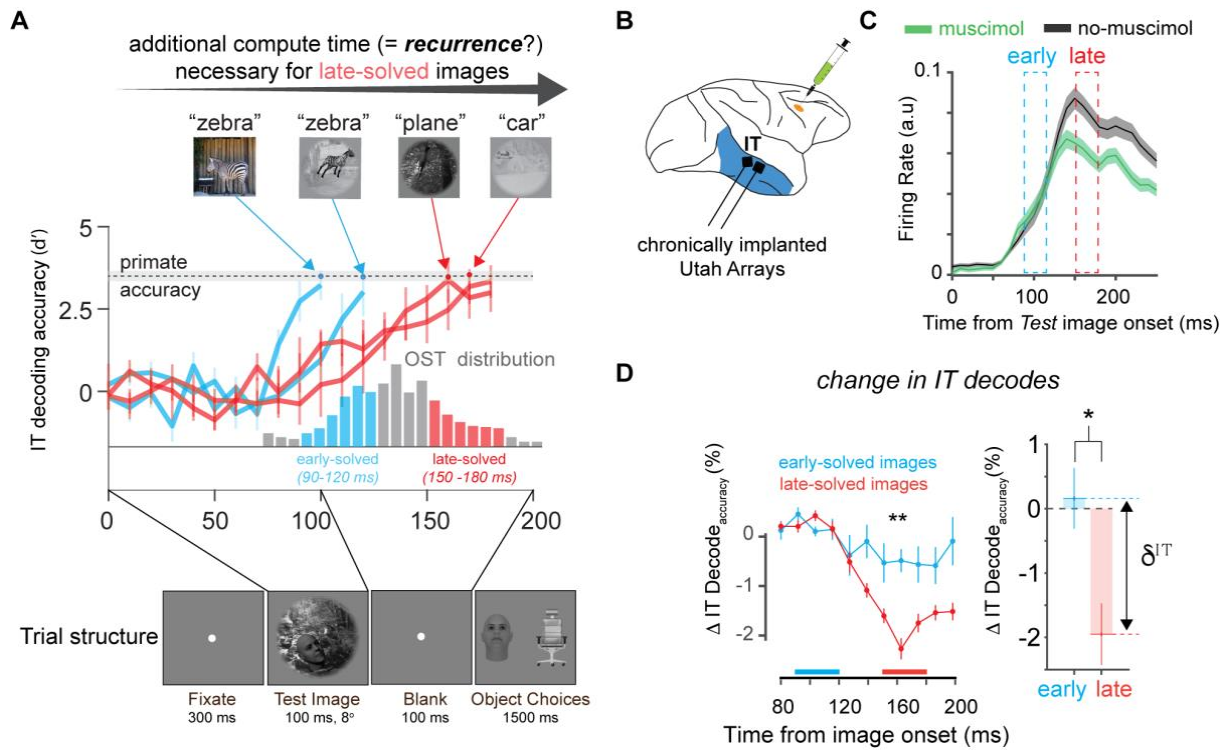

**Figure 2. Static images evoke reliable, behaviorally relevant dynamics in inferior temporal cortex driven by top-down modulation. A.** Time-resolved decoding of object identity from IT responses to briefly flashed static images reveals systematic temporal structure. Some images are resolved rapidly (*early-solved*, ≈90–120 ms; blue), whereas others require later IT activity to reach primate-level accuracy (*late-solved*, ≈150–180 ms; red). These late-solved images dominate the object solution time (OST; time at which the IT decode accuracy matches the monkey's behavioral accuracy) distribution, demonstrating that even static stimuli engage extended neural dynamics rather than a purely feedforward sweep. **B.** Experimental approach combining chronically implanted Utah arrays in IT with reversible pharmacological inactivation of ventrolateral prefrontal cortex (vlPFC), a major source of top-down input to IT. **C.** IT population responses with (green) and without (black) vlPFC inactivation. Early stimulus-evoked activity is largely preserved, whereas late-phase IT responses are selectively attenuated, indicating a specific disruption of feedback-dependent dynamics. **D.** Consequences for object decoding: inactivation has minimal impact on early-solved images but selectively impairs late-phase IT decodes for late-solved images, producing a pronounced late-epoch decoding deficit (ΔIT). Together, these results provide causal evidence that reliable temporal dynamics in IT, even for static images, depend on top-down recurrent interactions and are necessary for resolving challenging object representations. This figure was adapted from [12].

## 3. Dynamics Driven by Dynamic Stimuli

Dynamic visual stimuli provide a more ecological probe of ventral-stream computation. In natural vision, objects move, change shape, and interact with their surroundings, requiring the visual system to integrate information across time rather than across static snapshots. Studying ventral-stream responses to dynamic stimuli, therefore, allows us to test whether the intrinsic dynamics observed for static images generalize to continuous visual input, or whether new computational mechanisms are recruited. Neurons throughout V4 and IT integrate information over time, track features across motion, and generate expectations about future states [13].

Bigelow et al. [35] asked whether long-range motion computation arises entirely within the dorsal stream or whether the ventral stream contributes significantly. They found that V4 neurons encoded motion direction and speed only when motion was defined by a coherent moving object and not by aperture-limited gratings. This result indicates that the VVS performs object-centered motion computations that go beyond the integration of low-level motion cues.

Ramezanpour et al. [13] extended this line of work by recording from macaque IT during short (300 ms) video sequences in which objects moved at different speeds and in different directions. They observed that object motion could be decoded more accurately from IT than from V4, and that microstimulation of IT impaired speed discrimination performance. These findings provide causal evidence that the VVS contains signals necessary for perceiving dynamic object properties and that these signals are integrated with object identity representations.

## 3.1 Distributed circuit contributions to dynamic stimuli processing

Explaining these dynamic responses requires considering interactions across multiple pathways (Figure 3A). First, ventral and dorsal streams interact closely [36]. Motion-selective regions such as MT and MST provide inputs to the VVS [17,37], supplying information about speed [38], direction [39], and motion boundaries [40] that can support dynamic identity updates and object tracking. Second, higher-order areas, such as the prefrontal cortex [12] and hippocampus [41], influence VVS dynamics through signals of expectation, memory, and prediction. Thus, the dynamic responses observed in the VVS likely arise from coordinated operations across multiple brain areas.

## 3.2 Computational models of dynamic vision: from video recognition to object tracking

Similar to how convolutional networks from computer vision became some of the most brain-aligned models of feedforward VVS processing, current video-based neural networks (see example in Figure 3B), such as ConvLSTMs [42], 3D convolutional networks [43], two-stream architectures [44], and transformers [45], learn spatiotemporal features that support action recognition, dynamic scene understanding, and motion analysis. Models for object tracking and segmentation, including Siamese networks, space-time architectures, and more recent transformer-based improvements [46,47], maintain object identity over time despite motion. Although developed for engineering purposes, many of these systems embody computational strategies that resemble aspects of primate dynamic vision, such as integrating information across frames, maintaining continuity of object identity, and predicting future positions. Supporting this perspective, Tang et al. [48] showed that video-based networks predict human cortical responses during natural video viewing, and Dunnhofer et al. [49] demonstrated that these models provide better explanations (compared to feedforward models) of macaque IT dynamics in response to object motion.

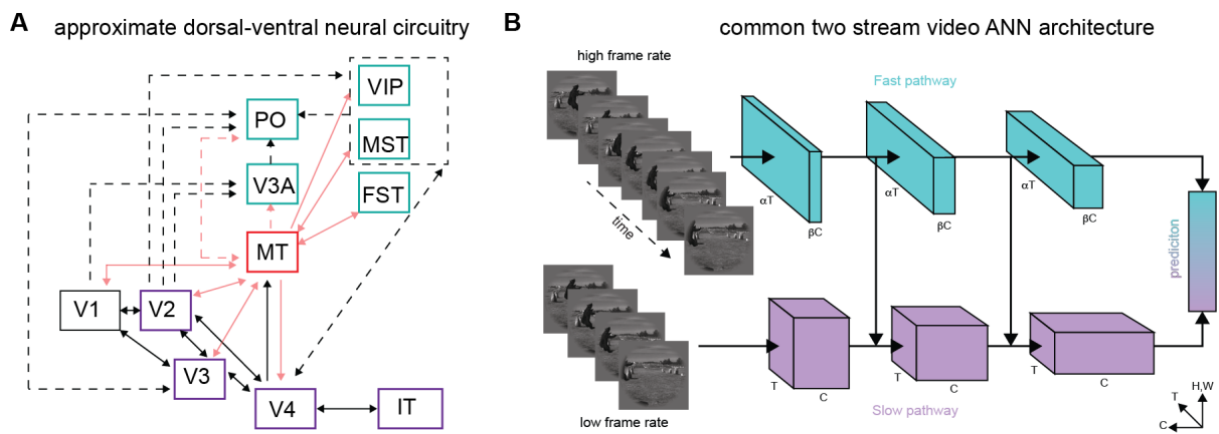

**Figure 3. Modeling dynamic vision requires targeted circuit-level experiments. A. Approximate dorsal–ventral neural circuitry in the primate visual system.** Early visual areas (V1–V3) project to both ventral-stream regions (V4, IT), supporting object recognition, and dorsal-stream regions involved in motion and spatial processing. Area MT occupies a central position in this network, serving as a major hub of interconnections with dense feedforward, feedback, and lateral interactions that link dorsal and ventral pathways (including projections to V4, V3A, MST, FST, VIP, and PO). Through these cross-stream and top-down interactions, MT provides a powerful route by which motion and temporal signals can influence ventral-stream representations, contributing to reliable neural dynamics even for static or briefly presented images. *Adapted from classic primate visual pathway schematics (Felleman & Van Essen, 1991), with simplifications.* **B. Common two-stream video ANN architecture.** Motivated by the separation of motion-sensitive and form-sensitive processing, two-stream video models (e.g., slowfast ANNs [44]) process high-frame-rate input through a fast pathway and low-frame-rate input through a slow pathway, with periodic information exchange across layers. While these architectures capture some stimulus-driven temporal structure, they abstract away the hub-like role of areas such as MT and the rich recurrent, multi-area interactions that shape dynamics in biological visual systems.

### 3.3. Divergent strategies for motion processing in brains and models

Video ANNs capture several early and mid-level dynamic computations in the VVS. However, they also reveal clear differences. Many high-performing video models rely heavily on appearance-based strategies [50] and perform poorly (Figure 4) when object identity information is minimized [51]. By contrast, primate VVS responses can generalize to appearance-free motion from their link to object shape and form during motion perception. Moreover, most video ANNs lack biologically grounded recurrence, multi-area coupling, or realistic temporal heterogeneity. They often struggle with late-phase dynamics, task- and state-dependent modulation, and top-down effects known to influence IT and V4.

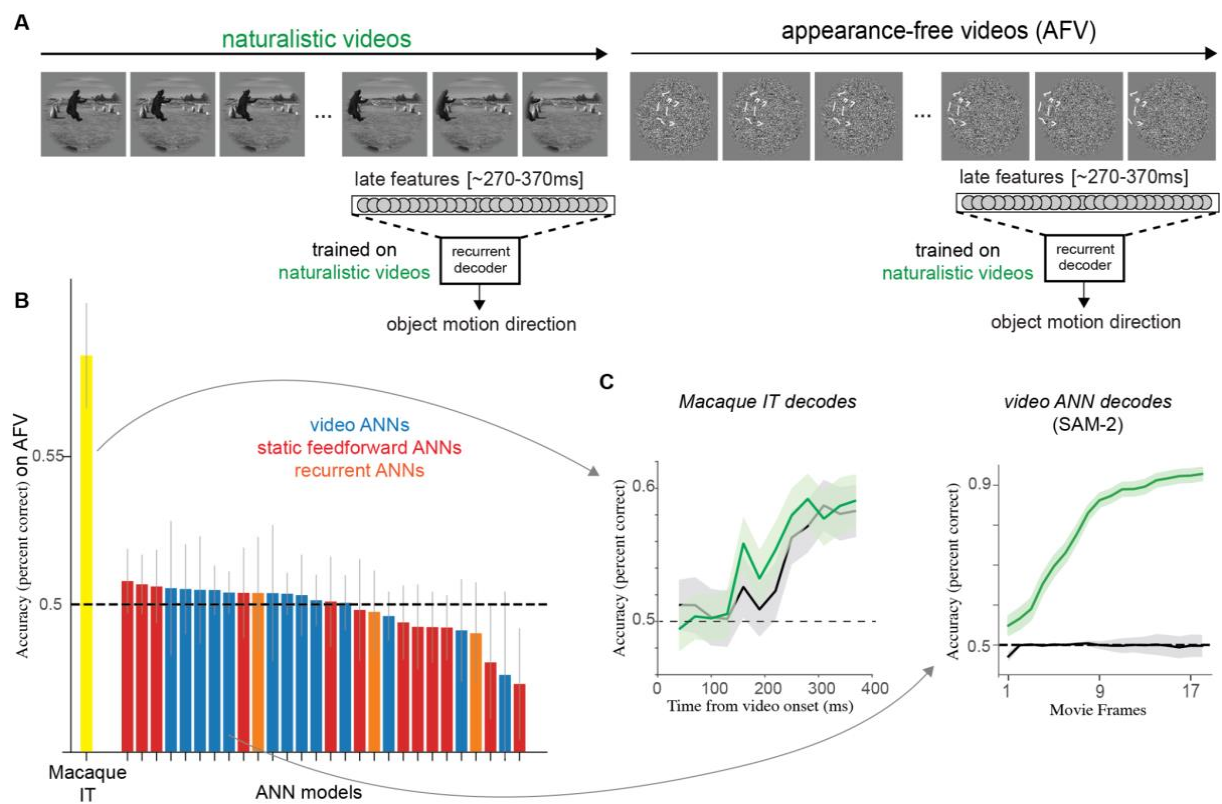

**Figure 4. Generalization of motion signals reveals divergent dynamic strategies in macaque IT and video ANNs. A.** Object motion direction decoding from late neural features (≈270–370 ms) extracted during viewing of naturalistic videos (left) and appearance-free videos (AFV; right). Decoders are trained on naturalistic videos and tested either on held-out naturalistic videos or on AFVs, which

preserve motion statistics while removing stable appearance cues. **B.** Decoding accuracy on AFVs across systems. Macaque IT (yellow) reliably supports motion direction decoding above chance despite the absence of appearance information. In contrast, most artificial neural network (ANN) models—including static feedforward (red), recurrent (orange), and video-based architectures (blue) fail to generalize to AFVs, performing at a chance level (chance = 0.5). **C.** Time-resolved decoding reveals that macaque IT representations progressively accumulate motion information over the course of the video. In contrast, video ANN representations (SAM-2 [52], shown here) achieve high accuracy only when appearance cues are present and collapse to chance when tested on AFVs.

### 3.4. Principles for closing the gap between video ANNs and VVS dynamics

Future models of dynamic vision should incorporate objectives aligned with natural behavior, including motion prediction, trajectory inference, and neural time-series alignment. Architectures should include recurrent loops that connect the ventral, dorsal, and prefrontal regions. Finally, training data should include naturalistic motion statistics, ego-motion, and complex object interactions to enable models to develop dynamic representations that more closely align with those observed in primates. These improvements represent key steps toward biologically grounded models of dynamic perception.

## 4. Dynamics from Dynamic Sensing: The Role of Eye Movements

### 4.1 Active vision as a dynamic input generator

Animals move their eyes to foveate objects of interest, using sequences of fixations and saccades that continually transform spatial structure into dynamic retinal input. Consequently, the VVS must encode the currently fixated object while coordinating with broader brain networks to guide subsequent eye movements [53,54]. During natural vision, the brain must handle continual retinal shifts, integrate information across fixations, and disentangle sensory signals from self-generated eye-movement dynamics; challenges largely absent from static-image paradigms.

### 4.2 Empirical evidence: VVS responses during exploration

Previous work has shown that VVS responses during free viewing differ from those measured under controlled fixation. Sheinberg and Logothetis [55] showed that IT neurons preserve object selectivity during natural visual search and can respond prior to fixation when an object is the goal of an upcoming saccade, linking IT activity to behavior during active vision. Consistent with this, pre-saccadic modulation in the intermediate visual cortex predicts forthcoming eye movements, suggesting a retinotopic salience signal shaped by attention[56]. Large-scale free-viewing recordings revealed that VVS responses remain predominantly gaze-dependent, with limited evidence for robust trans-saccadic feature persistence [57], while responses can nonetheless evolve across repeated fixations in an experience-dependent manner [58]. Together, these findings support the view that ventral visual representations are strongly fixation-locked but dynamically modulated by attention (see [59] for a review), temporal context, and behavioral goals during active vision [60].

## 4.3 Modeling active vision

Early computational accounts of active vision cast perception as a sequential sampling problem in which eye movements determine information acquisition over time. Bayesian active-sensing frameworks formalized fixation selection as information maximization, showing that human scan paths during visual search approximate optimal information-seeking policies [61,62], while sequential hypothesis-testing models framed object recognition as a series of disambiguating fixations [63,64]. Biologically inspired models then emphasized recurrent perception–action loops, coupling ventral-stream–like representations with oculomotor control to support invariant recognition during free viewing (e.g., ARTSCAN and related architectures [65]). With the rise of deep learning, reinforcement-learning–based attention models jointly learned fixation policies and perceptual representations from foveated inputs, enforcing biologically motivated sampling constraints [66,67]. More recent work focuses on predicting full scanpath rather than isolated fixations. Recurrent deep models generate temporally structured fixation sequences that incorporate history dependence and constraints such as inhibition of return [68–70], and benchmarks highlight scanpath modeling as a central goal for active vision [71]. In parallel, gaze has been used as a supervisory signal to reveal systematic mismatches between feedforward ANNs and human visual strategies [72]. Finally, event-based and transformer-inspired approaches extend active vision into the temporal domain at scale: event-camera pipelines learn eye-movement dynamics from high-resolution temporal streams [73], while emerging transformer models reinterpret saccadic sampling through scalable attention mechanisms, pointing toward a convergence between biological active vision and modern architectures.

## 4.4 Key insight

Visual neural dynamics are fundamentally shaped by active sensing: eye movements restructure visual input, inducing variability and transients across the ventral stream that static, feedforward models miss. Despite advances in active vision modeling, a key gap remains in linking these frameworks to electrophysiological measurements from primate VVS during natural viewing. Bridging this gap will require jointly modeling eye movements, task demands, and fixation-locked neural responses to more faithfully capture primate visual computation.

## 5. Beyond Representational Dynamics: The Need for Circuit-Level Models

Physiological recordings show that VVS responses reflect dynamical circuit mechanisms including interactions between excitatory and inhibitory populations [74,75], structured noise correlations [76,77], transient synchrony [78,79], and recurrent exchanges across cortical areas [9,28,80] that actively shape how representations evolve over time. These features rarely appear explicitly in current models and are not captured by static representational metrics, yet they are fundamental to how the system computes. As experimental paradigms increasingly probe the VVS under dynamic and naturalistic conditions [13,81,82], incorporating circuit-level structure will be essential for explaining the mechanisms that generate VVS dynamics. A key dynamic property of VVS circuits is the presence of time-varying interactions among neurons, including structured noise correlations [76] and synchrony [78]. These coordinated fluctuations evolve rapidly after stimulus onset [83], and influence coding efficiency, temporal integration, and downstream propagation [77,84,85]. Yet most models do not explicitly represent these population-level patterns (but see [86,87]), partly because benchmarks emphasize mean firing rates rather than temporal or population-level structure, and because most models do not generate trial-to-trial variability. Incorporating dynamical circuit mechanisms such as coordinated activity across neurons, neural noise, or shared temporal structure [86–88], is essential for capturing the population dynamics characteristic of the VVS.

Another missing ingredient is cell-type–specific circuitry. VVS circuits comprise excitatory pyramidal neurons and multiple classes of inhibitory interneurons [90,91], which differ in latency, integration, and connectivity [92,93]. These cell-type differences contribute to temporal phenomena, including gain control, normalization, suppression, and selective modulation. ANN models typically align more closely with excitatory than inhibitory responses [74], suggesting they might approximate excitatory-driven computations better than inhibitory-specific computations. Because ANN units do not obey biological constraints such as Dale's law [94] (but see [95]) or interneuron-like modulatory dynamics, they often fail to reproduce cell-type–dependent temporal patterns. Models that incorporate distinct neuronal types and connectivity may offer a more mechanistic account of VVS dynamics.

## 6. Conclusion

Across the evidence reviewed here, a clear conclusion emerges: the ventral visual stream is inherently dynamic, even when visual input is static. Early feedforward responses provide only an initial snapshot, while later activity reflects transformations shaped by recurrent circuitry, top-down influences, cross-area interactions, and behavioral state. Dynamic stimuli and active sensing further amplify this temporal structure, continuously reshaping visual representations. Existing artificial neural networks capture aspects of early processing but fail to reproduce the multi-area, multi-timescale dynamics that characterize primate vision. Progress will require models that move beyond static representations to explicitly incorporate the circuit mechanisms and temporal objectives that generate visual dynamics in natural behavior.

## References

1. Hung CP, Kreiman G, Poggio T, DiCarlo JJ: **Fast Readout of Object Identity from Macaque Inferior Temporal Cortex**. *Science* 2005, **310**:863–866.

2. Majaj NJ, Hong H, Solomon EA, DiCarlo JJ: **Simple Learned Weighted Sums of Inferior Temporal Neuronal Firing Rates Accurately Predict Human Core Object Recognition Performance**. *J Neurosci* 2015, **35**:13402–13418.

3. Kar K, DiCarlo JJ: **The Quest for an Integrated Set of Neural Mechanisms Underlying Object Recognition in Primates**. *Annual Review of Vision Science* 2024, **10**:91–121.

4. Kriegeskorte N: **Representational similarity analysis – connecting the branches of systems neuroscience**. *Front Sys Neurosci* 2008, doi:10.3389/neuro.06.004.2008.

5. Kornblith S, Norouzi M, Lee H, Hinton G: **Similarity of Neural Network Representations Revisited**. 2019, doi:10.48550/arXiv.1905.00414.

6. Yamins DLK, Hong H, Cadieu CF, Solomon EA, Seibert D, DiCarlo JJ: **Performance-optimized hierarchical models predict neural responses in higher visual cortex**. *Proc Natl Acad Sci USA* 2014, **111**:8619–8624.

7. Cadena SA, Denfield GH, Walker EY, Gatys LA, Tolias AS, Bethge M, Ecker AS: **Deep convolutional models improve predictions of macaque V1 responses to natural images**. *PLoS Comput Biol* 2019, **15**:e1006897.

8. Bashivan P, Kar K, DiCarlo JJ: **Neural population control via deep image synthesis**. *Science* 2019, **364**:eaav9436.

9. Kar K, Kubilius J, Schmidt K, Issa EB, DiCarlo JJ: **Evidence that recurrent circuits are critical to the ventral stream's execution of core object recognition behavior**. *Nat Neurosci* 2019, **22**:974–983.

10. Sugase Y, Yamane S, Ueno S, Kawano K: **Global and fine information coded by single neurons in the temporal visual cortex**. *Nature* 1999, **400**:869–873.

11*. Shi Y, Bi D, Hesse JK, Lanfranchi FF, Chen S, Tsao DY: **Rapid, concerted switching of the neural code in inferotemporal cortex**. *bioRxiv* 2023, doi:10.1101/2023.12.06.570341.

* Reveals rapid temporal code-switching in IT cortex, demonstrating that object representations dynamically reorganize across tens of milliseconds, challenging static notions of ventral-stream coding.

12. Kar K, DiCarlo JJ: **Fast Recurrent Processing via Ventrolateral Prefrontal Cortex Is Needed by the Primate Ventral Stream for Robust Core Visual Object Recognition**. *Neuron* 2021, **109**:164-176.e5.

13**. Ramezanpour H, Ilic F, Wildes RP, Kar K: **Object motion representation in the macaque ventral stream – a gateway to understanding the brain's intuitive physics engine**. 2024, doi:10.1101/2024.02.23.581841.

** Demonstrates that macaque IT cortex contains causal, behaviorally relevant motion signals, revealing a ventral-stream contribution to dynamic object processing.

14. Meyers EM, Freedman DJ, Kreiman G, Miller EK, Poggio T: **Dynamic population coding of category information in inferior temporal and prefrontal cortex**. *J Neurophysiol* 2008, **100**:1407–1419.

15. Wehrheim M, Alamooti ST, Ramezanpour H, Kar K: **Facial expression discrimination emerges from neural subspaces shared with detection and identity**. 2025, doi:10.1101/2025.08.25.672186.

16. Gilbert CD, Wiesel TN: **Columnar specificity of intrinsic horizontal and corticocortical connections in cat visual cortex**. *J Neurosci* 1989, **9**:2432–2442.

17. Felleman DJ, Van Essen DC: **Distributed Hierarchical Processing in the Primate Cerebral Cortex**. *Cerebral Cortex* 1991, **1**:1–47.

18. Stettler DD, Das A, Bennett J, Gilbert CD: **Lateral connectivity and contextual interactions in macaque primary visual cortex**. *Neuron* 2002, **36**:739–750.

19. Markov NT, Vezoli J, Chameau P, Falchier A, Quilodran R, Huissoud C, Lamy C, Misery P, Giroud P, Ullman S, et al.: **Anatomy of hierarchy: feedforward and feedback pathways in macaque visual cortex**. *J Comp Neurol* 2014, **522**:225–259.

20. Rockland KS, Pandya DN: **Laminar origins and terminations of cortical connections of the occipital lobe in the rhesus monkey**. *Brain Res* 1979, **179**:3–20.

21. Self MW, van Kerkoerle T, Supèr H, Roelfsema PR: **Distinct roles of the cortical layers of area V1 in figure-ground segregation**. *Curr Biol* 2013, **23**:2121–2129.

22. Amaral DG, Price JL, Pitkänen A, Carmichael ST: **Anatomical organization of the primate amygdaloid complex**. In *The Amygdala: Neurobiological Aspects of Emotion, Memory, and Mental Dysfunction*. . New York : Wiley-Liss; 1992.

23. Freese JL, Amaral DG: **Synaptic organization of projections from the amygdala to visual cortical areas TE and V1 in the macaque monkey**. *J Comp Neurol* 2006, **496**:655–667.

24. Suzuki WA, Amaral DG: **Perirhinal and parahippocampal cortices of the macaque monkey: cortical afferents**. *J Comp Neurol* 1994, **350**:497–533.

25. Arcaro MJ, Pinsk MA, Kastner S: **The Anatomical and Functional Organization of the Human Visual Pulvinar**. *J Neurosci* 2015, **35**:9848–9871.

26. Saalmann YB, Pinsk MA, Wang L, Li X, Kastner S: **The Pulvinar Regulates Information Transmission Between Cortical Areas Based on Attention Demands**. *Science* 2012, **337**:753–756.

27. Schrimpf M, Kubilius J, Hong H, Majaj NJ, Rajalingham R, Issa EB, Kar K, Bashivan P, Prescott-Roy J, Geiger F, et al.: *Brain-Score: Which Artificial Neural Network for Object Recognition is most Brain-Like?* Neuroscience; 2018.

28. Kietzmann TC, Spoerer CJ, Sörensen LKA, Cichy RM, Hauk O, Kriegeskorte N: **Recurrence is required to capture the representational dynamics of the human visual system**. *Proc Natl Acad Sci U S A* 2019, **116**:21854–21863.

29. Kubilius J, Schrimpf M, Kar K, Rajalingham R, Hong H, Majaj N, Issa E, Bashivan P, Prescott-Roy J, Schmidt K, et al.: **Brain-Like Object Recognition with High-Performing Shallow Recurrent ANNs**. In *Advances in Neural Information Processing Systems*. Edited by Wallach H, Larochelle H, Beygelzimer A, Alché-Buc F d', Fox E, Garnett R. Curran Associates, Inc.; 2019.

30. Lotter W, Kreiman G, Cox D: **A neural network trained for prediction mimics diverse features of biological neurons and perception**. *Nature Machine Intelligence* 2020, **2**:210–219.

31. Nayebi A, Bear D, Kubilius J, Kar K, Ganguli S, Sussillo D, DiCarlo JJ, Yamins DLK: **Task-Driven Convolutional Recurrent Models of the Visual System**. 2018, doi:10.48550/ARXIV.1807.00053.

32. Tang H, Schrimpf M, Lotter W, Moerman C, Paredes A, Ortega Caro J, Hardesty W, Cox D, Kreiman G: **Recurrent computations for visual pattern completion**. *Proceedings of the National Academy of Sciences of the United States of America* 2018, **115**:8835–8840.

33*. de Vries IEJ, Wurm MF: **Predictive neural representations of naturalistic dynamic input**. *Nat Commun* 2023, **14**:3858.

\* Shows that cortical responses reflect predictive representations of unfolding dynamic stimuli, emphasizing that temporal structure is an essential component of visual coding.

34*. Ostrow M, Eisen AJ, Kozachkov L, Fiete IR: **Beyond Geometry: Comparing the Temporal Structure of Computation in Neural Circuits with Dynamical Similarity Analysis**. 2023.

* Introduces principled tools for comparing neural and model dynamics beyond static representational geometry, enabling time-resolved evaluation of computational mechanisms.

35. Bigelow A, Kim T, Namima T, Bair W, Pasupathy A: **Dissociation in neuronal encoding of object versus surface motion in the primate brain**. *Current Biology* 2023, **33**:711-719.e5.

36. Milner AD: **How do the two visual streams interact with each other?** *Exp Brain Res* 2017, **235**:1297–1308.

37. Van Essen DC, Maunsell JHR, Bixby JL: **The middle temporal visual area in the macaque: Myeloarchitecture, connections, functional properties and topographic organization**. *J of Comparative Neurology* 1981, **199**:293–326.

38. Krekelberg B, Van Wezel RJA, Albright TD: **Interactions between Speed and Contrast Tuning in the Middle Temporal Area: Implications for the Neural Code for Speed**. *J Neurosci* 2006, **26**:8988–8998.

39. Albright TD: **Direction and orientation selectivity of neurons in visual area MT of the macaque**. *Journal of Neurophysiology* 1984, **52**:1106–1130.

40. Born RT, Bradley DC: **Structure and function of visual area MT**. *Annu Rev Neurosci* 2005, **28**:157–189.

41. Bonnen T, Yamins DLK, Wagner AD: **When the ventral visual stream is not enough: A deep learning account of medial temporal lobe involvement in perception**. *Neuron* 2021, **109**:2755-2766.e6.

42. Donahue J, Hendricks LA, Guadarrama S, Rohrbach M, Venugopalan S, Darrell T, Saenko K: **Long-term recurrent convolutional networks for visual recognition and description**. In *2015 IEEE Conference on Computer Vision and Pattern Recognition (CVPR)*. . IEEE; 2015:2625–2634.

43. Carreira J, Zisserman A: **Quo Vadis, Action Recognition? A New Model and the Kinetics Dataset**. In *2017 IEEE Conference on Computer Vision and Pattern Recognition (CVPR)*. . IEEE; 2017:4724–4733.

44. Feichtenhofer C, Fan H, Malik J, He K: **SlowFast Networks for Video Recognition**. In *2019 IEEE/CVF International Conference on Computer Vision (ICCV)*. . IEEE; 2019:6201–6210.

45. Bertasius G, Wang H, Torresani L: **Is Space-Time Attention All You Need for Video Understanding?** 2021, doi:10.48550/ARXIV.2102.05095.

46. Bertinetto L, Valmadre J, Henriques JF, Vedaldi A, Torr PHS: **Fully-Convolutional Siamese Networks for Object Tracking**. 2021,

47. Dunnhofer M, Furnari A, Farinella GM, Micheloni C: **Visual Object Tracking in First Person Vision**. *Int J Comput Vis* 2023, **131**:259–283.

48*. Tang Y, Gokce A, Al-Karkari KJ, Yamins D, Schrimpf M: **Diverse Perceptual Representations Across Visual Pathways Emerge from A Single Objective**. 2025, doi:10.1101/2025.07.22.664908.

*Demonstrates that training video models with unified objectives yields representations aligned across multiple visual pathways, offering a promising direction for modeling distributed cortical dynamics.

49. Dunnhofer M, Micheloni C, Kar K: **Dynamic Object Processing in Macaque IT Cortex: Temporal Dynamics and Model Limitations**. *Journal of Vision* 2025, **25**:2723.

50. Ilic F, Pock T, Wildes RP: **Is Appearance Free Action Recognition Possible?** In *Computer Vision – ECCV 2022*. Edited by Avidan S, Brostow G, Cissé M, Farinella GM, Hassner T. Springer Nature Switzerland; 2022:156–173.

51**. Dunnhofer M, Micheloni C, Kar K: **Better, But Not Sufficient: Testing Video ANNs Against Macaque IT Dynamics**. 2026, doi:10.48550/arXiv.2601.03392.

** Shows that video-based ANN models better capture stimulus-driven IT dynamics than static models, but still fail to account for late-phase and task-dependent neural responses.

52. Ravi N, Gabeur V, Hu Y-T, Hu R, Ryali C, Ma T, Khedr H, Rädle R, Rolland C, Gustafson L, et al.: **SAM 2: Segment Anything in Images and Videos**. In *The Thirteenth International Conference on Learning Representations*. . [date unknown].

53. Itti L, Koch C: **Computational modelling of visual attention**. *Nat Rev Neurosci* 2001, **2**:194–203.

54. Desimone R, Duncan J: **Neural Mechanisms of Selective Visual Attention**. *Annual Review of Neuroscience* 1995, **18**:193–222.

55. Sheinberg DL, Logothetis NK: **Noticing Familiar Objects in Real World Scenes: The Role of Temporal Cortical Neurons in Natural Vision**. *J Neurosci* 2001, **21**:1340–1350.

56. Mazer JA, Gallant JL: **Goal-Related Activity in V4 during Free Viewing Visual Search**. *Neuron* 2003, **40**:1241–1250.

57**. Xiao W, Sharma S, Kreiman G, Livingstone MS: **Feature-selective responses in macaque visual cortex follow eye movements during natural vision**. *Nat Neurosci* 2024, **27**:1157–1166.

** Large-scale recordings during free viewing show that ventral-stream responses are strongly fixation-locked, providing crucial constraints for models of active visual processing.

58*. Yamane Y, Ito J, Joana C, Fujita I, Tamura H, Maldonado PE, Doya K, Grün S: **Neuronal Population Activity in Macaque Visual Cortices Dynamically Changes through Repeated Fixations in Active Free Viewing**. *eNeuro* 2023, **10**:ENEURO.0086-23.2023.

* Demonstrates experience-dependent evolution of VVS population activity across fixations, highlighting the importance of temporal context during active vision.

59. Ramezanpour H, Fallah M: **The role of temporal cortex in the control of attention**. *Current Research in Neurobiology* 2022, **3**:100038.

60.	Ramezanpour H, Thier P: **Decoding of the other's focus of attention by a temporal cortex module**. *Proc Natl Acad Sci USA* 2020, **117**:2663–2670.

61.	Najemnik J, Geisler WS: **Optimal eye movement strategies in visual search**. *Nature* 2005, **434**:387–391.

62.	Yang SC-H, Lengyel M, Wolpert DM: **Active sensing in the categorization of visual patterns**. *eLife* 2016, **5**:e12215.

63.	Schill K, Umkehrer E, Beinlich S, Krieger G, Zetzsche C: **Knowledge-based scene analysis with saccadic eye movements**. In Edited by Rogowitz BE, Pappas TN. 1999:520–531.

64.	Rybak IA, Gusakova VI, Golovan AV, Podladchikova LN, Shevtsova NA: **A model of attention-guided visual perception and recognition**. *Vision Research* 1998, **38**:2387–2400.

65.	Fazl A, Grossberg S, Mingolla E: **View-invariant object category learning, recognition, and search: How spatial and object attention are coordinated using surface-based attentional shrouds**. *Cognitive Psychology* 2009, **58**:1–48.

66.	Larochelle H, Hinton GE: **Learning to combine foveal glimpses with a third-order Boltzmann machine**. In *Advances in Neural Information Processing Systems*. Edited by Lafferty J, Williams C, Shawe-Taylor J, Zemel R, Culotta A. Curran Associates, Inc.; 2010.

67.	Mnih V, Heess N, Graves A, Kavukcuoglu K: **Recurrent Models of Visual Attention**. In *Advances in Neural Information Processing Systems*. Edited by Ghahramani Z, Welling M, Cortes C, Lawrence N, Weinberger KQ. Curran Associates, Inc.; 2014.

68.	Wang X, Zhao X, Ren J: **A New Type of Eye Movement Model Based on Recurrent Neural Networks for Simulating the Gaze Behavior of Human Reading**. *Complexity* 2019, **2019**:8641074.

69.	Tang Y, Su J: **Eye Movement Prediction Based on Adaptive BP Neural Network**. *Scientific Programming* 2021, **2021**:1–9.

70.	Chen Z, Sun W: **Scanpath Prediction for Visual Attention using IOR-ROI LSTM**. In *Proceedings of the Twenty-Seventh International Joint Conference on Artificial Intelligence*. . International Joint Conferences on Artificial Intelligence Organization; 2018:642–648.

71.	Li J, Watters N, Yingting, Wang, Sohn H, Jazayeri M: **Modeling Human Eye Movements with Neural Networks in a Maze-Solving Task**. 2022, doi:10.48550/ARXIV.2212.10367.

72.	Van Dyck LE, Denzler SJ, Gruber WR: **Guiding visual attention in deep convolutional neural networks based on human eye movements**. *Front Neurosci* 2022, **16**:975639.

73.*	Seth C, Naiken D, Lin K: **A deep learning approach to track eye movements based on events**. 2025, doi:10.48550/ARXIV.2508.04827.

* Demonstrates event-based modeling of eye-movement dynamics at high temporal resolution, bridging active sensing, temporal vision, and scalable machine-learning approaches.

74. Sanghavi S, Kar K: *Distinct roles of putative excitatory and inhibitory neurons in the macaque inferior temporal cortex in core object recognition behavior*. Neuroscience; 2023.

75. Tamura H, Kaneko H, Kawasaki K, Fujita I: **Presumed Inhibitory Neurons in the Macaque Inferior Temporal Cortex: Visual Response Properties and Functional Interactions With Adjacent Neurons**. *Journal of Neurophysiology* 2004, **91**:2782–2796.

76. Cohen MR, Kohn A: **Measuring and interpreting neuronal correlations**. *Nat Neurosci* 2011, **14**:811–819.

77. Averbeck BB, Latham PE, Pouget A: **Neural correlations, population coding and computation**. *Nat Rev Neurosci* 2006, **7**:358–366.

78. Fries P: **A mechanism for cognitive dynamics: neuronal communication through neuronal coherence**. *Trends in Cognitive Sciences* 2005, **9**:474–480.

79. Singer W: **Neuronal Synchrony: A Versatile Code for the Definition of Relations?** *Neuron* 1999, **24**:49–65.

80. Lamme VAF, Roelfsema PR: **The distinct modes of vision offered by feedforward and recurrent processing**. *Trends in Neurosciences* 2000, **23**:571–579.

81. Hasson U, Malach R, Heeger DJ: **Reliability of cortical activity during natural stimulation**. *Trends in Cognitive Sciences* 2010, **14**:40–48.

82. Huk A, Bonnen K, He BJ: **Beyond Trial-Based Paradigms: Continuous Behavior, Ongoing Neural Activity, and Natural Stimuli**. *J Neurosci* 2018, **38**:7551–7558.

83. Smith MA, Kohn A: **Spatial and Temporal Scales of Neuronal Correlation in Primary Visual Cortex**. *J Neurosci* 2008, **28**:12591–12603.

84. Abbott LF, Dayan P: **The Effect of Correlated Variability on the Accuracy of a Population Code**. *Neural Computation* 1999, **11**:91–101.

85. Zohary E, Shadlen MN, Newsome WT: **Correlated neuronal discharge rate and its implications for psychophysical performance**. *Nature* 1994, **370**:140–143.

86. Reichert DP, Serre T: **Neuronal Synchrony in Complex-Valued Deep Networks**. 2013, doi:10.48550/ARXIV.1312.6115.

87*. Muzellec S, Alamia A, Serre T, VanRullen R: **Enhancing deep neural networks through complex-valued representations and Kuramoto synchronization dynamics**. 2025, doi:10.48550/arXiv.2502.21077.

* Introduces synchronization-based dynamics into neural networks, providing a concrete computational framework for modeling population-level temporal structure observed in cortical circuits.

88.	Behrmann M, Zemel RS, Mozer MC: **Object-based attention and occlusion: Evidence from normal participants and a computational model.** *Journal of Experimental Psychology: Human Perception and Performance* 1998, **24**:1011–1036.

89.	Ravishankar Rao A, Cecchi GA, Peck CC, Kozloski JR: **Unsupervised Segmentation With Dynamical Units**. *IEEE Trans Neural Netw* 2008, **19**:168–182.

90.	Tremblay R, Lee S, Rudy B: **GABAergic Interneurons in the Neocortex: From Cellular Properties to Circuits**. *Neuron* 2016, **91**:260–292.

91.	Markram H, Toledo-Rodriguez M, Wang Y, Gupta A, Silberberg G, Wu C: **Interneurons of the neocortical inhibitory system**. *Nat Rev Neurosci* 2004, **5**:793–807.

92.	The Petilla Interneuron Nomenclature Group (PING): **Petilla terminology: nomenclature of features of GABAergic interneurons of the cerebral cortex**. *Nat Rev Neurosci* 2008, **9**:557–568.

93.	Isaacson JS, Scanziani M: **How Inhibition Shapes Cortical Activity**. *Neuron* 2011, **72**:231–243.

94.	Eccles JC: **From electrical to chemical transmission in the central nervous system: The closing address of the Sir Henry Dale Centennial Symposium Cambridge, 19 September 1975**. *Notes Rec R Soc Lond* 1976, **30**:219–230.

95.	Li P, Cornford J, Ghosh A, Richards B: *Learning better with Dale's Law: A Spectral Perspective*. Neuroscience; 2023.